\documentclass[aip,reprint,showpacs,twocolumn,superscriptaddress,floatfix]{revtex4-1}
\usepackage{epsfig}
\usepackage[T1]{fontenc}
\usepackage[utf8]{inputenc}
\usepackage{amsmath}
\usepackage{amssymb}
\usepackage{amsfonts}
\usepackage{mathptmx}
\usepackage{dcolumn}
\usepackage{eucal}
\usepackage{bm}
\usepackage{color}
\usepackage[colorlinks,linkcolor=blue,citecolor=blue]{hyperref}
\pdfpageattr {/Group << /S /Transparency /I true /CS /DeviceRGB>>}
\begin{document}
\title{A ballistic two-dimensional-electron-gas Andreev interferometer}

\author{M. Amado}
\email{mario.amadomontero@sns.it}
\affiliation{NEST, Istituto Nanoscienze-CNR and Scuola Normale Superiore, I-56127 Pisa, Italy}

\author{A. Fornieri}
\affiliation{NEST, Istituto Nanoscienze-CNR and Scuola Normale Superiore, I-56127 Pisa, Italy}

\author{G. Biasiol}
\affiliation{CNR-IOM, Laboratorio TASC, Area Science Park, I-34149 Trieste, Italy}

\author{L. Sorba}
\affiliation{NEST, Istituto Nanoscienze-CNR and Scuola Normale Superiore, I-56127 Pisa, Italy}

\author{F. Giazotto}
\email{f.giazotto@sns.it}
\affiliation{NEST, Istituto Nanoscienze-CNR and Scuola Normale Superiore, I-56127 Pisa, Italy}

%%%%%%%%%%%%%%%%%%%%%%%%%%%%%%%%%%%%%%%%%%%%%%%%%%%%%%%%


%%%%%%%%%%%%%%%%%%%%%%%%%%%%%%%%%%%%%%%%%%%%%%%%%%%%%%%%%%%%
%%%%%%%%%%%%%%%%%%%%%   ABSTRACT         %%%%%%%%%%%%%%%%%%%
%%%%%%%%%%%%%%%%%%%%%   ABSTRACT         %%%%%%%%%%%%%%%%%%%
%%%%%%%%%%%%%%%%%%%%%   ABSTRACT         %%%%%%%%%%%%%%%%%%%
%%%%%%%%%%%%%%%%%%%%%   ABSTRACT         %%%%%%%%%%%%%%%%%%%
%%%%%%%%%%%%%%%%%%%%%   ABSTRACT         %%%%%%%%%%%%%%%%%%%
%%%%%%%%%%%%%%%%%%%%%%%%%%%%%%%%%%%%%%%%%%%%%%%%%%%%%%%%%%%%

\begin{abstract}
We report the realization and investigation of a ballistic Andreev interferometer based on an InAs two dimensional electron gas coupled to a superconducting Nb loop. We observe strong magnetic modulations in the voltage drop across the device due to quasiparticle interference within the weak-link. The interferometer exhibits flux noise down to $\sim 80\, \mu\Phi_0/\sqrt{\textrm{Hz}}$, and a robust behavior in temperature with voltage oscillations surviving up to $\sim7\,$K. Besides this remarkable performance, the device represents a crucial first step for the realization of a fully-tunable ballistic superconducting magnetometer and embodies a potential advanced platform for the investigation of Majorana bound states, non-local entanglement of Cooper pairs, as well as the manipulation and control of spin triplet correlations.

\end{abstract}

%\pacs{73.23.Ad;	  % Ballistic transport
%      73.21.Fg;      % Quantum wells
%      85.25.Dq;	  % Superconducting quantum interference devices (SQUIDs)}

\maketitle

%%%%%%%%%%%%%%%%%%%%%%%%%%%%%%%%%%%%%%%%%%%%%%%%%%%%%%%%%%%%
%%%%%%%%%%%%%%%%%%%%   INTRODUCTION   %%%%%%%%%%%%%%%%%%%%%%
%%%%%%%%%%%%%%%%%%%%   INTRODUCTION   %%%%%%%%%%%%%%%%%%%%%%
%%%%%%%%%%%%%%%%%%%%   INTRODUCTION   %%%%%%%%%%%%%%%%%%%%%%
%%%%%%%%%%%%%%%%%%%%   INTRODUCTION   %%%%%%%%%%%%%%%%%%%%%%
%%%%%%%%%%%%%%%%%%%%   INTRODUCTION   %%%%%%%%%%%%%%%%%%%%%%
%%%%%%%%%%%%%%%%%%%%%%%%%%%%%%%%%%%%%%%%%%%%%%%%%%%%%%%%%%%%

\maketitle

The combination of ballistic two-dimensional-electron-gases (2DEGs) and superconductors (S) may represent a key tool for the investigation of Majorana bound states~\cite{Alicea,Beenakker-Review,Qi-RMP,Sau-PRL}, the generation of solid-state entanglers~\cite{Lefloch-PRL,Lefloch-arxiv} as well as spin triplet superconductivity~\cite{Bergeret-RMP,Eschrig-PToday}. Up to now, most of the related experimental works have been focused on the study of hybrid devices based on diffusive semiconductor nanowires~\cite{Mourik-Sci,Das-NatPhys}, normal metals~\cite{Lefloch-PRL,Lefloch-arxiv} and ferromagnets~\cite{Bergeret-RMP}. Nevertheless, the structure flexibility and the reduced influence of disorder candidate S-2DEG systems as a promising platform to reach the complete understanding of these effects~\cite{Aguado-PRL,Valentini-PRB}. One of the first steps toward this scope is the fabrication of a ballistic S-2DEG interferometer, in which the phase of quasiparticles is controlled by an external magnetic field and the number of conducting channels can be tailored at will thanks to the presence of side gates. The realization of this kind of device must fulfill two essential conditions: (i) a Schottky barrier-free normal region-superconductor interface is necessary to couple the superconductor with the weak link leading to a robust proximity effect, and (ii) the geometrical dimensions of the 2DEG-region must lie below the electron elastic mean free path ensuring the ballistic transport of quasiparticles in the normal region. In the last few years, the major efforts have been put on InAs and InAs/AlSb 2DEG-based hybrid structures. These materials exhibit a larger \emph{g}-factor and spin-orbit coupling in comparison to those existing in In$_{0.75}$Ga$_{0.25}$As semiconductor alloys~\cite{Capotondi-vacuum,Desrat-PRB2,Carillo-PE,Deon-PRB}, being therefore more attractive for the implementation of a topological non-trivial phase~\cite{Alicea,Beenakker-Review,Qi-RMP}. Moreover, the interplay between spin-orbit coupling and spin-splitting (due to the application of an external magnetic field) may lead to an advanced manipulation and control of the triplet superconducting correlations~\cite{Bergeret-PRL,Bergeret-PRB} with potential application in spintronics.
\begin{figure}[t!]
\centerline{\includegraphics[width=\columnwidth,clip=]{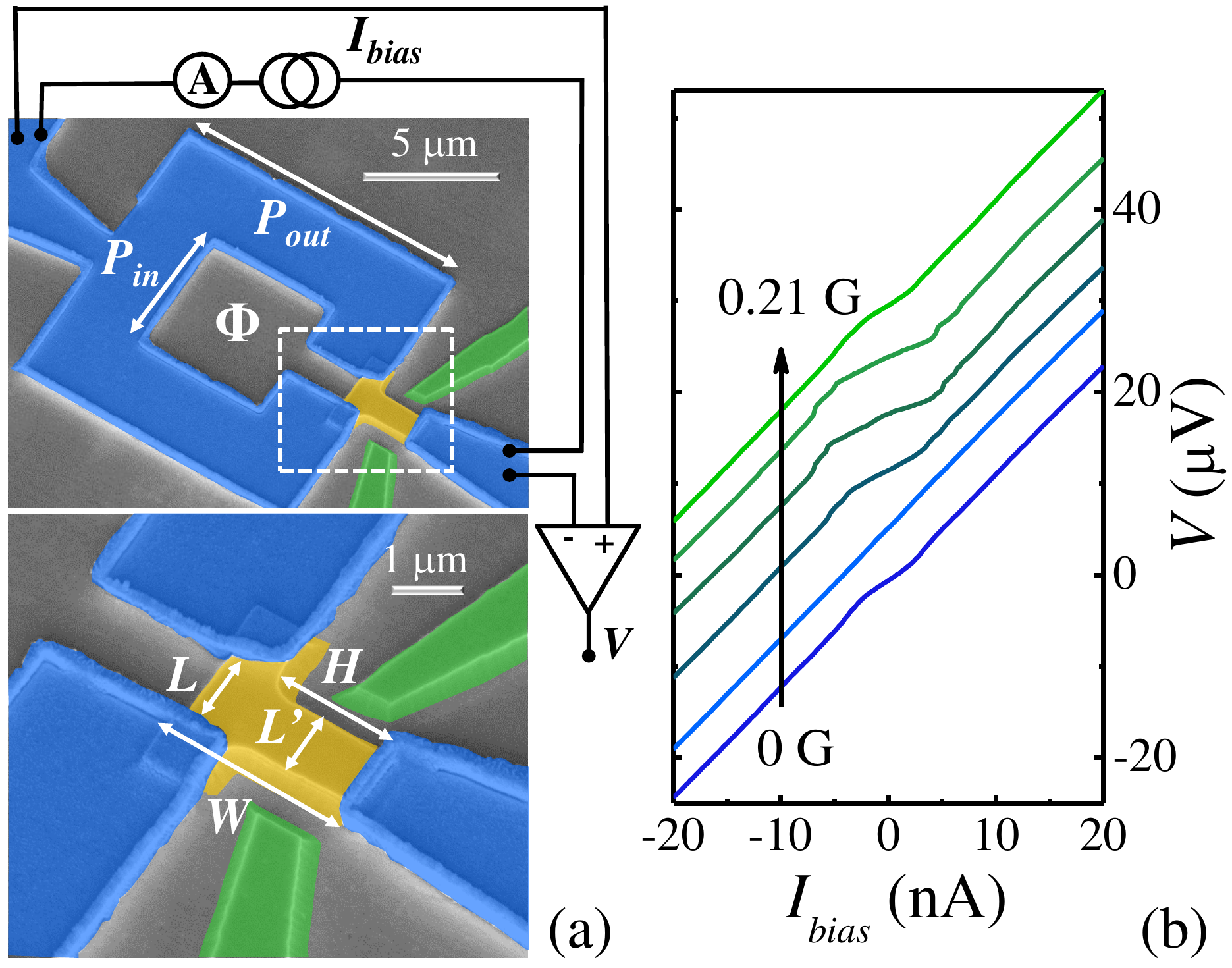}}
\caption{(a) Pseudo-color scanning electron micrograph of a typical device consisting of an InAs 2DEG T-shaped island (yellow) interrupting a Nb superconducting loop (blue) and connected to a superconducting probe (also in blue) with two-etched side gates (green). An external magnetic flux ($\Phi$) is applied perpendicular to the 2DEG's plane. The blow-up shows a detail of the hybrid interferometer core. The interelectrode spacing in the upper end of T-shape 2DEG is $L\sim1.2\,\mu$m while the distance between the loop and the superconducting probe is $H\sim2\,\mu$m. The standard current-bias 4-wires set-up for the measurements appears sketched in the circuit. Panel (b) displays voltage vs. current ($V$ vs. $I_{bias}$) characteristics measured for some selected values of the external magnetic field at $10\,$mK. The curves have been vertically shifted by $3\,\mu$V for clarity.}\label{sample}
\end{figure}
While the early generation of magnetic-flux controlled interferometers consisted of a normal metal (N) region connected to ring-shaped superconductors ~\cite{Petrashov-PRL,Petrashov-JETP,Courtois-PRL,Charlat-PRL,Antonov-PhysicaC,Belzig-PRB}, the first realization of a 2DEG-based quasiparticle interferometer was reported by Dimoulas \emph{et al.}~\cite{Dimoulas-PRL}, where the interplay between Josephson coupling and quasiparticle interference was investigated. Subsequent works were devoted to the study of the charge transport at the 2DEG/S interfaces in the diffusive~\cite{Hartog-PRL,Hartog2-PRL} and ballistic limits~\cite{Morpurgo-PRL,Morpurgo2-PRL} of ring-shaped~\cite{Hartog-PRL}, microcavities~\cite{Morpurgo-PRL,Morpurgo2-PRL} and quantum-point-contact-like (QPC-like) structures~\cite{Hartog2-PRL}, exploiting all of them these promising Schottky barrier-free semiconductors as weak links.

In this Letter we present the fabrication and the investigation of a mesoscopic Andreev interferometer. In contrast with the most recent devices based on proximized diffusive InAs nanowires~\cite{Spathis-NNT,Giazotto-Natphys2,Chang-PRL}, a ballistic T-shaped 2DEG InAs mesa is the basis of our weak-link. The latter is connected to a ring-shaped niobium (Nb) electrode and to a third superconducting probe, which is used to measure the voltage drop across the device when threading the loop with an external magnetic field. The interferometer offers a remarkable accuracy as a magnetometer with a flux noise of $\sim 80 \mu\Phi_0/\sqrt{\textrm{Hz}}$ and a robust performance up to a temperature of $\sim7\,$K. Furthermore, in contrast to N/S interferometers, additional side gates can be used to modify the carrier density, thereby tuning the conductance of the 2DEG-region and the response of the device.

The InAs quantum well-based (QW-based) heterostructure was grown by means of molecular beam epitaxy and is based on a GaAs (001) substrate on top of which a series of $50$-nm-thick In$_{1-x}$Al$_x$As layers was deposited (the concentration in Al varies from $x=0.85$ in the first layer to $x=0.25$ in the latter one). A $4$-nm-thick InAs QW is then interposed between two $5.5$-nm-thick In$_{0.75}$Ga$_{0.25}$As layers and asymmetric In$_{0.75}$Al$_{0.25}$As barriers~\cite{Capotondi-Thin}. The sheet electron density $n\simeq3.74\times10^{11}$ cm$^{-2}$, the mobility $\mu\simeq2\times10^5\,$cm$^2$/Vs and the electron mass $m^*\simeq0.03\,$m$_e$ were extracted from low-temperature Shubnikov-de-Haas oscillations measurements. We can therefore estimate the elastic mean free path of the 2DEG resulting in $l_0\simeq2~\mu$m.

The fabrication of the Andreev interferometer required a sequence of mutually aligned steps of electron beam lithography (EBL) as previously reported~\cite{Fornieri-NNT,Amado-PRB}. The ohmic contacts for the side gates were obtained in the first EBL step while the second self-aligned lithography was performed to define the mesa region of the 2DEG, i.e., the T-shaped central island of the interferometer and the two etched side gates. To this end, a negative resist bilayer was spin coated on the surface of the sample and served us as the mask defining the 2DEG-region. The surface of the heterostructure was then attacked by means of a chemical wet etching in a H$_2$O:H$_2$SO$_4$:H$_2$O$_2$ solution. The final mesa has a typical total width $W\sim2.7~\mu$m while the width of the vertical strip of the T-shaped 2DEG is $L'\sim900\,$nm. The superconducting parts of the interferometer were designed by the last step of EBL.
Prior to the sputter deposition of the $200$-nm-thick Nb film, the interfaces were cleaned from undesired oxide layer with a dip into a HF:H$_2$O solution and a low-energy Ar$^{+}$ milling in the sputtering chamber.

A typical device is displayed in Fig.~\ref{sample}(a), where the superconducting leads appear in blue, the weak-link in yellow and the side gates in green. The inter-electrode spacing between the arms of the ring is $L\sim1.2\,\mu$m, while the third superconducting probe is connected to the 2DEG at a distance $H\sim2\,\mu$m from the loop. The loop has an inner length side $P_{in}\sim5\,\mu$m whereas the outer one is $P_{out}\sim10\,\mu$m. Finally, the side gates are placed $\sim900\,$nm away from the mesa.

The hybrid interferometers were characterized in a filtered dilution refrigerator down to $10\,$mK. The structure was biased by a current $I_{bias}$ whereas the voltage drop $V$ across the mesa has been registered via a room-temperature differential preamplifier [see Fig.~\ref{sample}(a)]. Typical $V$ vs. $I_{bias}$ characteristics obtained from the 4-wires set-up are shown in Fig.~\ref{sample}(b). The curves (taken at $10\,$mK and vertically offset by $3\,\mu$V) are related to different external magnetic fields $B$ ranging from $0$ to $0.21\,$G. As it can be noticed, small changes in $B$ have a major impact on the interferometer's response, which will be the subject of our analysis.
\begin{figure}[t!]
\centerline{\includegraphics[width=\columnwidth,clip=]{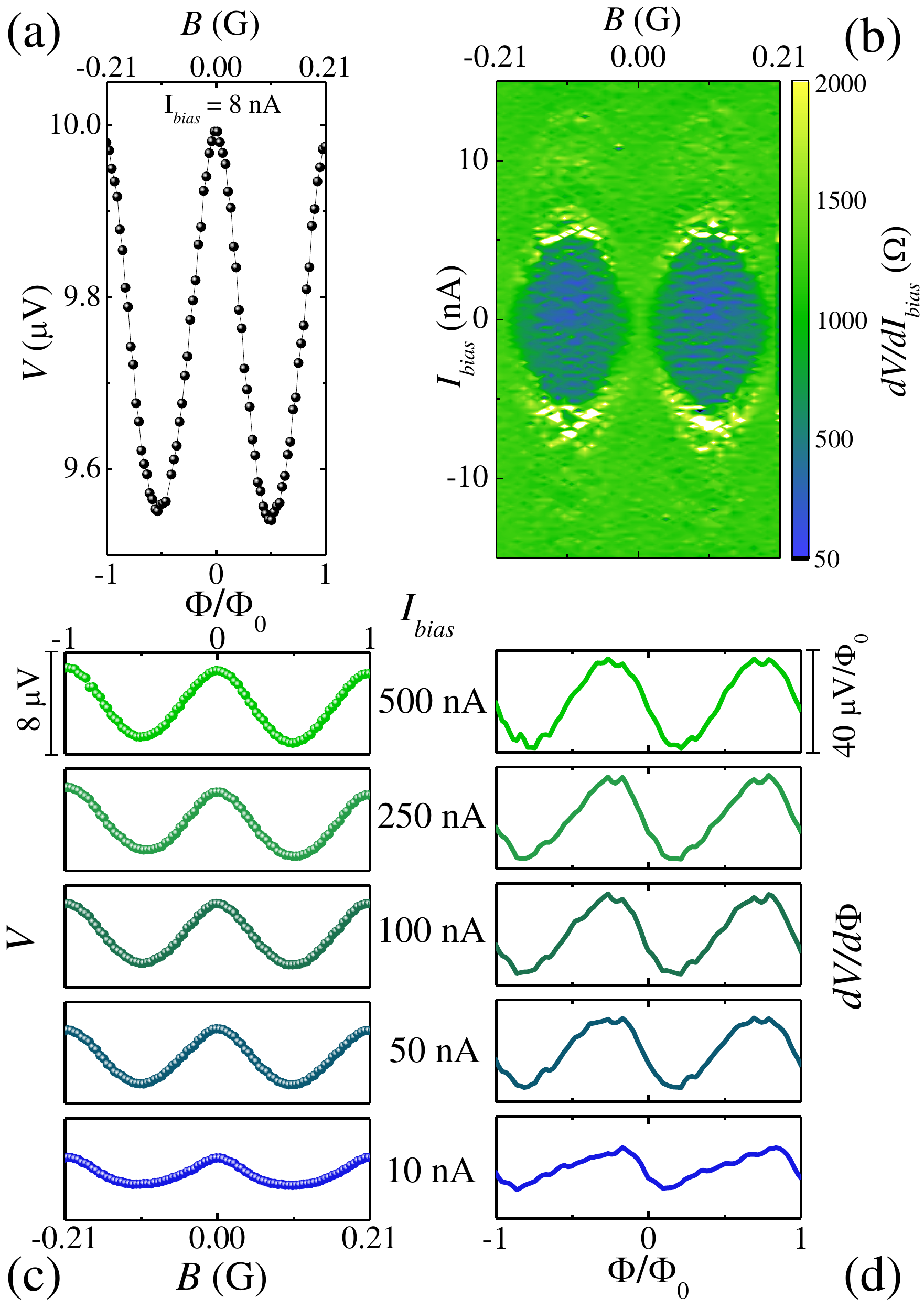}}
\caption{(a) Evolution of $V$ as a function of the external magnetic field $B$ and the magnetic flux $\Phi$ for $I_{bias}=8\,$nA. (b) Color plot showing the dependence of the numerical differential resistance $dV/dI_{bias}$ on $I_{bias}$ and $B$. (c) $V$ vs. $B$ and $\Phi$ for a few selected values of $I_{bias}$. (d) Flux-to-voltage transfer function $dV/d\Phi$ vs. $\Phi$ for the same values of $I_{bias}$ considered in panel (c). All the curves have been measured at $10\,$mK.}\label{dVdI}
\end{figure}

Fig.~\ref{dVdI}(a) shows the evolution of $V$ as a function of $B$ for two complete periods measured at $10\,$mK and $I_{bias}=8\,$nA. The magnetic field dependence of the voltage drop across the device is intimately related to the changes in the phase difference along the 2DEG/S interfaces and represents the basis of the Andreev interferometer. Such changes in the acquired phase are transferred through the Andreev reflection process to the quasiparticles in the weak link which lead to the interference in $V$~\cite{Dimoulas-PRL,Pothier-PRL}. Since the superconducting's probe phase varies in a negligible way as a function of $B$, the modulation of $V$ is the consequence of the phase difference gained at the 2DEG/S boundaries on the interrupted Nb loop. Furthermore, we neglect the phase difference acquired along the superconducting arms of the Nb ring due to the low kinetic inductance of the superconductor in comparison to that of the 2DEG weak-link. As we can observe in Fig.~\ref{dVdI}(a), the period of the modulations is $\sim0.21\,$G which corresponds to an area $A\sim98\,\mu$m$^2$ extracted from $\Phi_0=A\times B$ (where $\Phi_0\simeq 2\times 10^{-15}\,$Wb is the superconducting flux quantum). Such value of the area corresponds to the one delimited by the outer perimeter of the Nb loop $A_{out}= P_{out}^2\sim100\,\mu$m$^2$ therefore confirming that the quasiparticle interference is mainly localized in the proximized 2DEG region within the ring interruption [see Fig.~\ref{sample}(a)].

The effect of the magnetic field is highlighted by the numerical differential resistance ($dV/dI_{bias}$) dependence on $I_{bias}$ and $B$ registered at the cryostat base temperature, shown in Fig.~\ref{dVdI}(b). The interference pattern is clearly visible for two complete periods with a noticeable blue-colored low-resistance central region at $|I_{bias}|\leq6\,$nA, indicating a precursor of Josephson current flowing through the weak-link.

Panel (c) of Fig.~\ref{dVdI} shows a relevant feature of the Andreev interferometer, i.e., the evolution of $V(B)$ for different values of $I_{bias}$ recorded at the cryostat base temperature. The amplitude of the $V(B)$ oscillations grows considerably by increasing $I_{bias}$, therefore enlarging their visibility and converging to a maximum value of the peak-to-peak amplitude of $\sim 8 \,\mu$V for $I_{bias}\geq600\,$nA, where we recover the normal state resistance of the sample $R_{N}\sim1.87\,$k$\Omega$. On the other hand, the visibility of the $dV/dI_{bias}$ oscillations decreases with increasing $I_{bias}$, as previously noticed~\cite{Hartog-PRL,Morpurgo-PRL}. For $I_{bias}=0$ the maximum peak to peak amplitude is $\sim900\,\Omega$ yielding a sizeable visibility of $\sim50\%$, about $25\%$ greater than that obtained in similar devices so far~\cite{Dimoulas-PRL}.
\begin{figure}[t!]
\centerline{\includegraphics[width=\columnwidth,clip=]{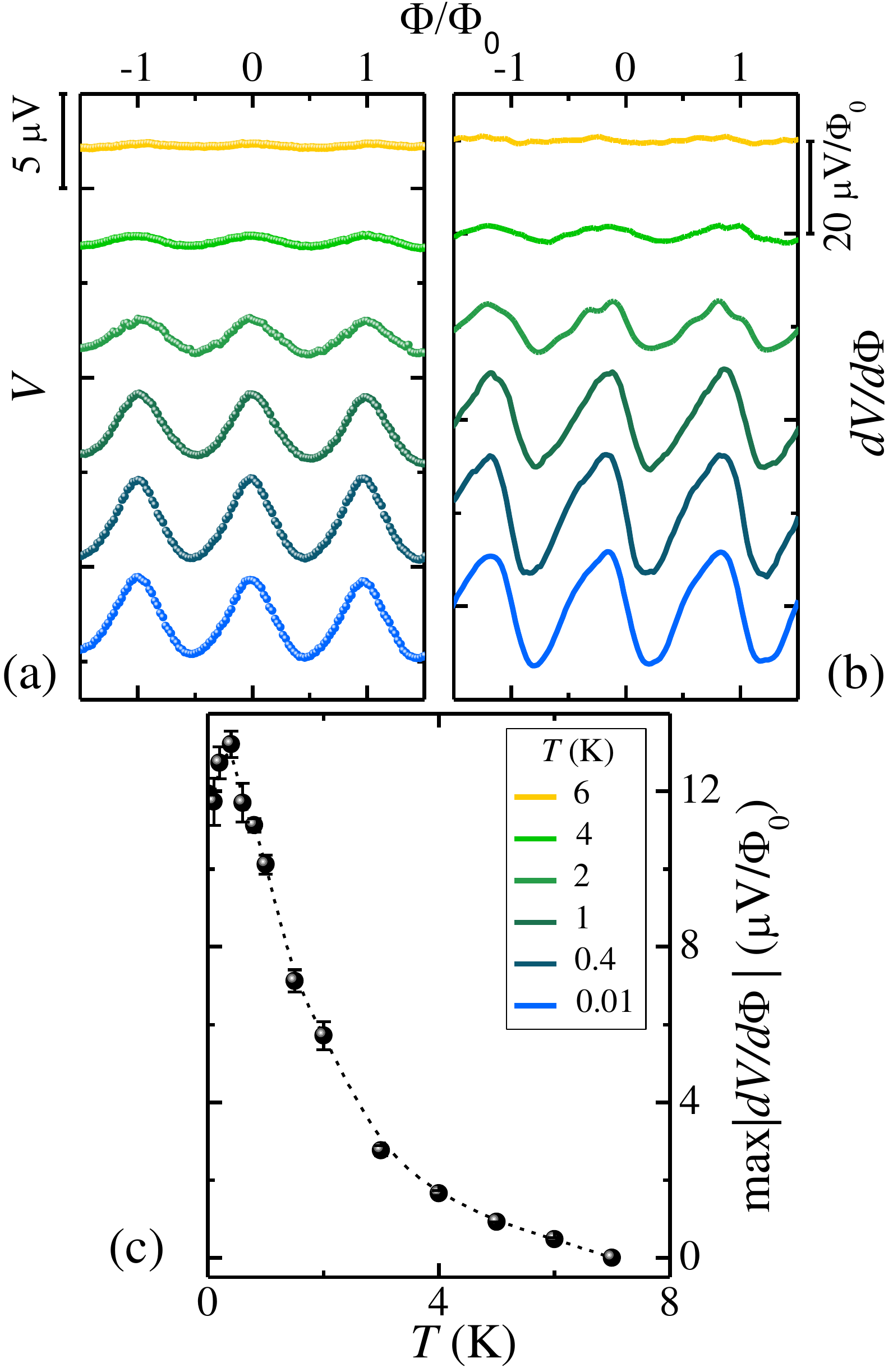}}
\caption{(a) $V$ vs. $\Phi$ measured at different temperatures $T$. (b) $dV/d\Phi$ vs. $\Phi$ at the same temperatures of panel (a). $T$ spans from $10\,$mK to $6\,$K and the curves have been vertically shifted by $5\,\mu$V in (a) and by $20\, \mu$V$/\Phi_0$ in (b) with an $I_{bias}=30\,$nA. (c) Evolution of the maximum of the flux-to-voltage transfer function max$|dV/d\Phi|$ vs. $T$ for $I_{bias}=30\,$nA. The short-dashed line is a guide to the eye.}\label{transfer-function}
\end{figure}
The flux-to-voltage transfer function ($dV/d\Phi$) vs. magnetic flux is shown in Fig.~\ref{dVdI}(d) for different values of $I_{bias}$. Its visibility grows with $I_{bias}$, faster at low values of the injection current and then saturating at $\sim25\,\mu$V$/\Phi_0$ for $I_{bias}\geq600\,$nA. We can therefore estimate the flux resolution of the interferometer from $\Phi_N=\sqrt{S_v}/$max$|dV/d\Phi|\sim80\,\mu\Phi_0/\sqrt{\textrm{Hz}}$, mainly determined by the noise of the room-temperature voltage preamplifier (with a typical $\sqrt{S_v}\sim2\,$nV$/\sqrt{\textrm{Hz}}$) and comparable with that measured in the first generation of metallic superconducting quantum interference proximity transistors~\cite{Giazotto-Natphys} (SQUIPTs).

Figure~\ref{transfer-function} displays the figures of merit of the interferometer as a function of temperature $T$. Panels (a) and (b) show the magnetic flux dependence of $V$ and $dV/d\Phi$ at a few selected values of $T$ and for $I_{bias}=30\,$nA. All the curves were vertically offset for clarity by $5\,\mu$V [panel (a)] and by $20\,\mu$V$/\Phi_0$ [panel (b)]. The amplitude of the oscillations decreases at larger $T$ due to the reduction of the coherence length in the 2DEG and of the Andreev reflection probability. Nevertheless, the $V(\Phi)$ oscillations remain visible well above the sub-Kelvin regime, surviving up to $\sim7\,$K. This is pointed out by Fig.~\ref{transfer-function}(c), which shows the maximum value of the transfer function (max$|dV/d\Phi|$) as a function of $T$ for $I_{bias}=30\,$nA. max$|dV/d\Phi|$ is barely modified within the sub-kelvin regime and approaches zero at larger temperatures. We emphasize that our device represents the first operational high-temperature Andreev interferometer in contrast to previous similar 2DEG-based systems that were exploited only in the mK-regime~\cite{Dimoulas-PRL,Hartog-PRL,Hartog2-PRL,Morpurgo-PRL}. Moreover, since the critical temperature of the Nb leads is $T_c\sim8\,$K~\cite{Fornieri-NNT}, the robustness of the performance of the interferometer confirms the high transparency of the 2DEG/S interfaces and makes our system attractive for its implementation as a hybrid magnetometer also at liquid $^4$He temperatures.

\begin{figure}[t!]
\centerline{\includegraphics[width=0.9\columnwidth,clip=]{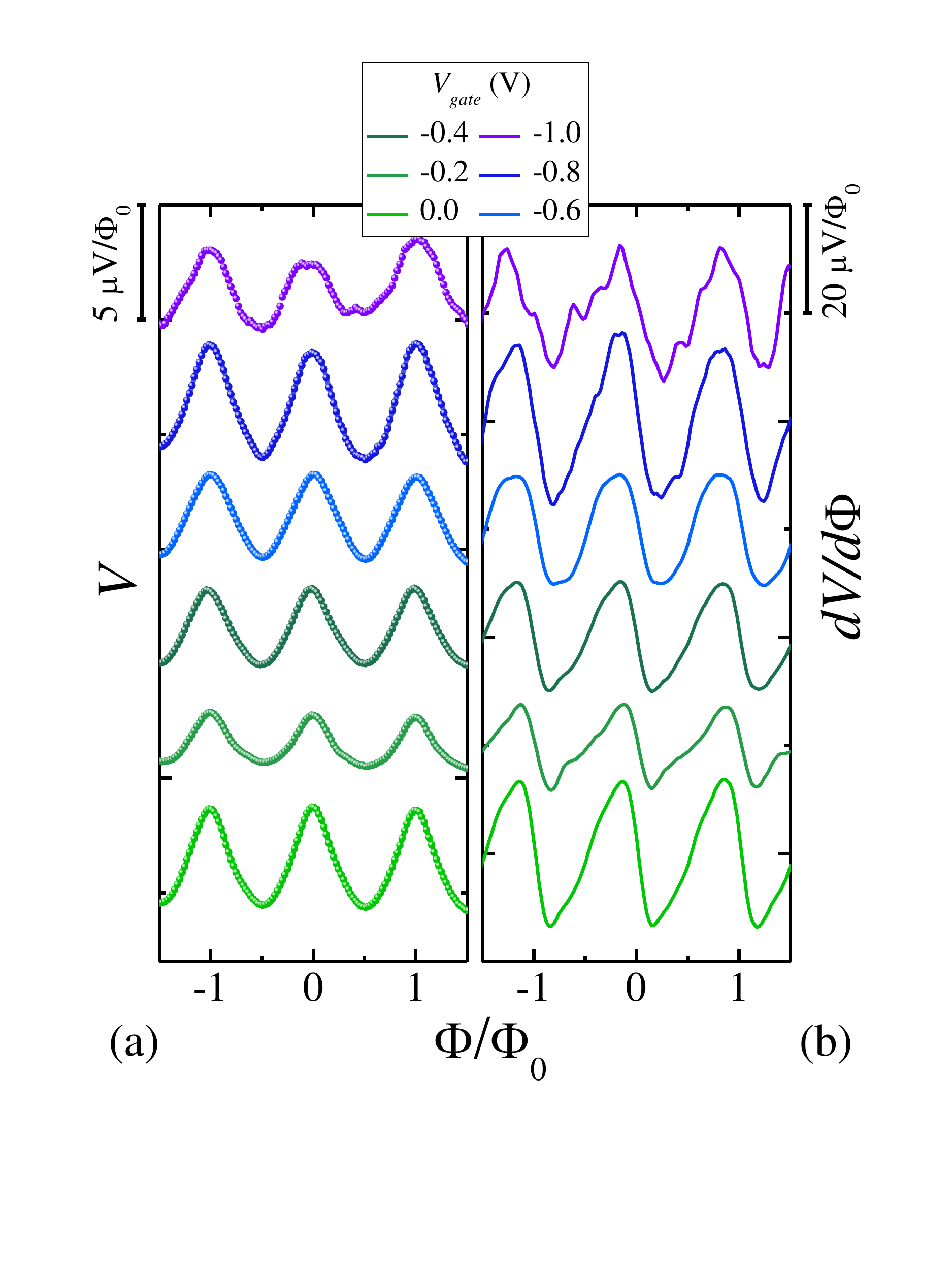}}
\caption{Lateral gate voltage ($V_{gate}$) dependence of $V$ [panel (a)] and $dV/d\Phi$ [panel (b)] vs. $\Phi$. $V_{gate}$ ranges from $0\,$V to $-1\,$V. All the curves have been vertically shifted by $20\, \mu$V$/\Phi_0$ and were recorded for $I_{bias}=30\,$nA at $10\,$mK.}\label{gate-effect}
\end{figure}

Finally, we introduce the effect of the side gates on the quasiparticle transport through the 2DEG. The normal state resistance of the interferometer varies when squeezing the QPC by biasing negatively the lateral side gates. While $V_{gate}\geq-1\,$V, $R_N$ varies smoothly arriving to a maximum value $R_N\sim4.8\,$k$\Omega$ but by outrunning such gate voltage the conductance in the system drops dramatically arriving to the pinch-off at $V_{gate}\sim-1.9\,$V. Fig.~\ref{gate-effect} shows the magnetic flux dependence of $V$ [panel (a)] and $dV/d\Phi$ [panel (b)] for different values of $V_{gate}$. The voltage spans from 0 to $-1\,$V with an injection current of $30\,$nA. The curves were recorded at $10\,$mK and have been been vertically offset by $5\,\mu$V [panel (a)] and $20\,\mu$V$/\Phi_0$ [panel (b)]. The typical shape in the $V(\Phi)$ and in the transfer function characteristics is maintained at low values of $|V_{gate}|$ but the response of the interferometer starts to be distorted near $V_{gate}\sim-1\,$V. At low values of $V_{gate}$, the maximum value of the transfer function of the interferometer is being modified by the external electric field with a non-monotonic behavior. On the other hand, by approaching the pinch-off regime, the voltage modulation loses its harmonic evolution against $\Phi$ and no traces of periodicity can be recovered. A possible explanation of such effect is the fact that the depopulation of the 2DEG might occur not only in the QPC-like region amid the lateral gates but in the whole mesa. Yet, the vicinity of the side gates to the 2DEG/S interface of the superconducting probe might also affect its transparency making more difficult to recover the magnetic modulation of $V$ at high $|V_{gate}|$.

In summary, we have reported the fabrication and investigation of a ballistic Andreev interferometer based on an InAs 2DEG strongly coupled to Nb leads. We measured a sizeable sensitivity to magnetic flux with a maximum value of the transfer function $dV/d\Phi\,\sim25\,\mu$V$/\Phi_0$ and a low flux noise down to $\sim80\,\mu\Phi_0/\sqrt{\textrm{Hz}}$. Our system can operate from the miliKelvin regime up to $T\sim7\,$K, therefore being promising for its implementation in sensitive magnetometry and in quantum circuits as, for instance, Cooper pairs entanglers~\cite{Lefloch-PRL,Lefloch-arxiv}. This device represents the first step towards the generation of highly-sensitive magnetometers, whose performance can be tuned \textit{in-situ} by controlling the transparency of the contact between the superconducting probe and the weak-link. In the low-transparency regime, power dissipation could be strongly limited~\cite{Giazotto-Natphys} and Andreev bound state spectroscopy might be also achieved~\cite{Pillet-NatPhys}. Finally, in presence of an appropriate magnetic field, the device could be an important tool for the investigation of topological superconductivity~\cite{Valentini-PRB} and long-range triplet superconducting correlations~\cite{Bergeret-PRL,Bergeret-PRB}.

We thank F. Carillo for his valuable advice on the fabrication process, as well as C. Altimiras and A. Ronzani for fruitful discussions. Partial financial support from the Marie Curie Initial Training Action (ITN) Q-NET 264034 and the Tuscany region through the project "TERASQUID" is acknowledged. The work of F.G. has been partially funded by the European Research Council under the European Union's Seventh Framework Programme (FP7/2007-2013)/ERC grant agreement No. 615187-COMANCHE.

\end{document}